# The Hypercube of Life: How Protein Stability Imposes Limits on Organism Complexity and Speed of Molecular Evolution.

Konstantin B. Zeldovich[1], Peiqiu Chen[1,2], Eugene I. Shakhnovich[1]

[1]Department of Chemistry and Chemical Biology,  [2]Department of Physics, Harvard University, 12 Oxford Street, Cambridge, MA 02138

**Classical population genetics a'priori assigns fitness to alleles without considering molecular or functional properties of proteins that these alleles encode. Here we study population dynamics in a model where fitness can be inferred from physical properties of proteins under a physiological assumption that loss of stability of any protein encoded by an essential gene confers a lethal phenotype. Accumulation of mutations in organisms containing $\Gamma$ genes can then be represented as diffusion within the $\Gamma$-dimensional hypercube with adsorbing boundaries which are determined, in each dimension, by loss of a protein's stability and, at higher stability, by lack of protein sequences. Solving the diffusion equation whose parameters are derived from the data on point mutations in proteins, we determine a universal distribution of protein stabilities, in agreement with existing data. The theory provides a fundamental relation between mutation rate, maximal genome size and thermodynamic response of proteins to point mutations. It establishes a universal speed limit on rate of molecular evolution by predicting that populations go extinct (via lethal mutagenesis) when mutation rate exceeds approximately 6 mutations per essential part of genome per replication for mesophilic organisms and 1-2 mutations per genome per replication for thermophilic ones. Further, our results suggest that in absence of error correction, modern RNA viruses and primordial genomes must necessarily be very short. Several RNA viruses function close to the evolutionary speed limit while error correction mechanisms used by DNA viruses and non-mutant strains of bacteria featuring various genome lengths and mutation rates have brought these organisms universally about 1000-fold below the natural speed limit.**

Phenomenological approaches to study molecular evolution, developed since pre-DNA era, assume selective advantage of certain alleles or the existence of a single fitness peak in the genome space [1]. While mathematical genetics approaches brought remarkable insights over the years, they lack a fundamental connection between the fitness (reproductive success) of organisms and molecular properties of proteins encoded by their genomes. On the other hand, our understanding of molecular basis of folding, stability and function of proteins has advanced significantly. In particular, statistical-mechanical studies provided key insights into sequence requirements for proteins to fold and be stable in their native, functional states [2]. Here, we develop an evolutionary model where the fitness of an organism can be, in principle, directly inferred from its DNA sequences using a protein folding model and a well-defined physiological assumption of genotype-phenotype relationship. This physiological assumption is motivated by recent experiments which showed that knockout of any essential gene confers a lethal phenotype to an organism [3] [4]. The number of such essential genes varies from organism to organism: all genes are essential in viruses while in bacteria essential genes can reach up to 1/3 of all genes. As a minimal functional requirement proteins have to be stable in their native conformations. Therefore, here we assume the following fundamental genotype-phenotype relationship: *in order for an organism to be viable all of its essential genes must encode (at least minimally) stable proteins*. While this requirement is certainly minimal, it is necessary, for essential genes, and universal. There have been conflicting opinions as to whether greater stability of proteins confers selective advantage or disadvantage or it is neutral [5-7]. Here we assume that protein stability is essentially a physiologically neutral trait insofar a protein possess sufficient stability to stay in the folded state [7,8]. However, proteins accumulate mutations over the course of evolution. While many mutations may be neutral with respect to protein stability [7] and some of them can be stabilizing, eventually accumulation of too may mutations will render the protein unstable, non-functional, and it will confer a lethal phenotype to its carrier organism.

Therefore, molecular evolution can be rendered as diffusion in space of protein stabilities. In order to describe the space in which such random walks occur we turn now to an elementary consideration of thermodynamics of protein folding.

In their native conformations, proteins possess low energy and low entropy while in the denatured (unfolded) state their energy and entropy are both high. Proteins unfold in a two-state manner at temperature $T_F$ when free energy of the native state $G_F$ equals to free energy of the unfolded state $G_U$ [9]:

$$G_F = E_F = G_U = E_U - k_B T_F S_U \qquad (1)$$

where $k_B$ $E_F, E_U, S_U$ are the Boltzmann constant, free energy, energy and entropy of folded and unfolded states respectively and we assumed here that entropy of the folded state is small[9,10]. Mutations affect mostly compact native state, by changing its energy $E_F$ [11,12]. A protein gets unfolded when its energy reaches value $E_{max}$ such that it loses stability at the temperature T at which the organism lives,

$$\Delta G = G_F(E_{max}) - G_U(E_U, S_U, T) = 0 \qquad (2)$$

or assuming the entropy of unfolding approximately constant for a standard size 100-amino-acid domain [9]

$$E_{max} = E_U - k_B T S_U \qquad (3)$$

(This is a simplified description, see Supplementary Information for a more complete analysis). Therefore if the energy of the native state of a mutant protein exceeds $E_{max}$, the protein becomes unstable and dysfunctional, causing death of an organism that carries its gene.(See Fig.1)

Due to the effect of sequence depletion, the range of the possible native energies of proteins is limited from below as well: there are simply no sequences that can deliver energies below a certain threshold energy $E_{min}$.[2,13,14] (see Supp Information for estimate of $E_{min}$).

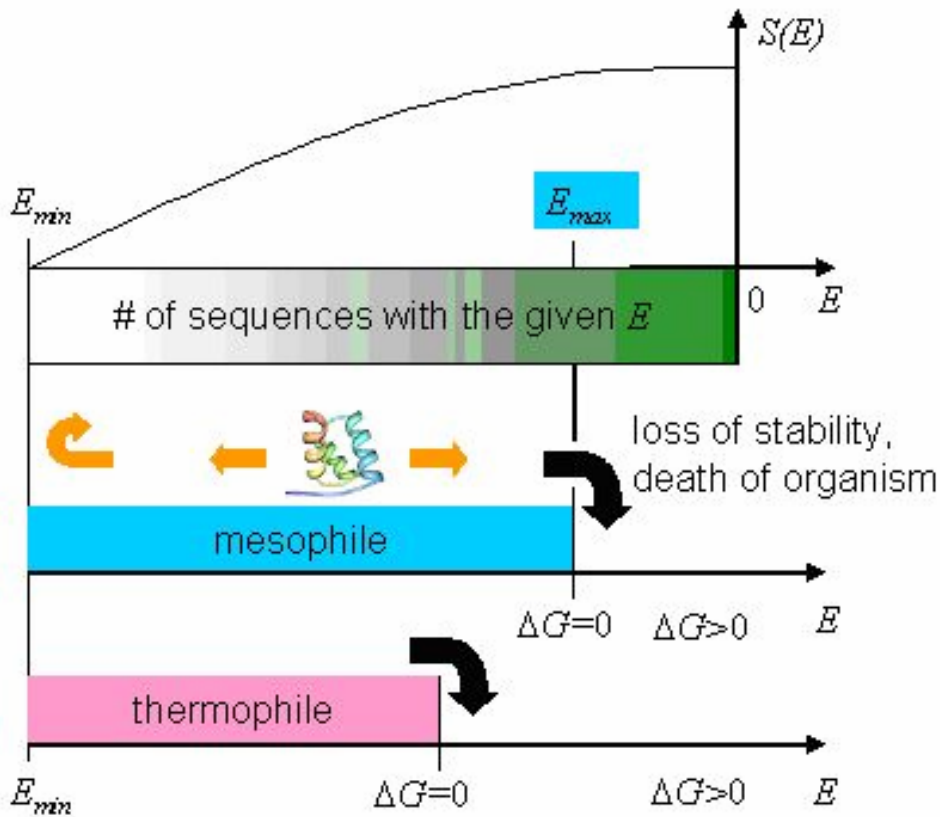

**Figure 1.** *A schematic representation of the model. Upper panel illustrates the effect of sequence depletion as energy of the native state of a protein decreases. The lower panels show accessible range of native state energies within which proteins can mutate. It is limited from below by $E_{min}$ - where complete sequence depletion occurs. From above this range is limited by $E_{max}$ when proteins become unstable conferring lethal phenotype to the organism carrying this protein's gene. The range of possible energies is broader for mesophilic organisms and narrows for thermophilic ones*

At any given time $t$, the genotype of an organism that carries $\Gamma$ genes can be fully characterized in this model by the native energies of the corresponding $\Gamma$ proteins, i.e. as a point $(E_1, E_2...E_\Gamma)$ in the $\Gamma$-dimensional space. We assume that changes of stability of different proteins upon mutations are independent and can be statistically described by the probability distribution $W(E_i^{WT} \to E_i^{mut}) = W(E_i^{WT}, E_i^{mut})$. *In vitro* experiments on mutant proteins provide a reasonable approximation for the shape of $W$ (see [11] and Supplementary Information). As each protein accumulates mutations in its sequence, its stability changes, and so is the position of the organism in the stability space $\vec{E}(t)$ (we use vector notation here to highlight the $\Gamma$-dimensional character of space of native energies of all proteins in a genome).

This model is equivalent to the following $\Gamma$-dimensional diffusion problem. Population consisting of N non-interacting organisms represents N independent random walkers each diffusing in $\Gamma$-dimensional space. Organisms replicate with rate b and there is natural, mutation-unrelated decay of genomes (death rate due to e.g. degradation of viral RNA) d. Introducing the joint (not normalized) distribution of genotypes in a population, i.e. *number* of organisms $P(\vec{E}, t)$ having genotype $\vec{E} = (E_1, E_2...E_\Gamma)$ at time t, we can write the replication-mutation balance diffusion equation

$$\frac{\partial P(E_1...E_i,..E_\Gamma, t)}{\partial t} = \frac{1}{2}(h^2 + D)m \sum_{i=1}^{\Gamma} \frac{\partial^2}{\partial E_i^2} P(E_1..E_i...E_\Gamma, t) + hm \sum_{i=1}^{\Gamma} \frac{\partial}{\partial E_i} P(E_1..E_i...E_\Gamma, t) + (b-d) P(E_1,..E_i..E_\Gamma, t)$$

(4)

(see Supp Info for derivation and detailed discussion of its applicability) The first two terms are due to the mutational flux of organisms to and from the vicinity of point $\vec{E}$, the third term corresponds to replication of organisms. m is a mutation rate *per gene*. Very importantly the equation (4) should be augmented by $2\Gamma$ boundary conditions $P(E_1..E_i....E_\Gamma) = 0$ for any $E_i > E_{max}$ and for any $E_i < E_{min}$ corresponding to organismal death and sequence depletion conditions for each gene as explained above.

Eq.(4) provides a deterministic description of evolution of distribution of genotypes. It is applicable in the regime when $Nm > 1$ [7]

We assume that organism replication rate $b$ does not depend on $\vec{E}$, so the fitness landscape is flat and there are only two phenotypes, the viable phenotype for $E_{min}<E_i<E_{max}$, $i=1\ldots\Gamma$, and a lethal phenotype otherwise.

As mutations in all proteins are independent, and boundary conditions are the same in each dimension (i.e. for each gene) one can write $P(\mathbf{E})=\prod_{i=1}^{\Gamma}p(E_i)$, separate the variables, and reduce the problem to a product of $\Gamma$ one-dimensional diffusion problems, for each gene:

$$\frac{\partial p}{\partial t} = \frac{b}{\Gamma}p + mh\frac{\partial p}{\partial E} + \frac{1}{2}m(h^2+D)\frac{\partial^2 p}{\partial E^2}, \quad p(E_{max})=0, \; p(E_{min})=0. \quad (5)$$

$$h = \int_{-\infty}^{\infty} W(E,E')(E-E')dE' \text{ and } h^2+D = \int_{-\infty}^{\infty} W(E,E')(E-E')^2 dE'$$

are mean and mean square (de)stabilization effect upon point mutations. The analysis of the data on point mutations available in the ProTherm database[15] and in [11] gives $h \approx 1(kcal/mol)$; $D \approx 3(kcal/mol)^2$.

The long-time solution of Eq.(4) has the form $P(E,t)=e^{\lambda t}P(E)$. The steady-state solution of the equation (5) is

$$p(E) = Ae^{-\frac{hE}{h^2+D}}\sin\left(\pi\frac{E-E_{min}}{E_{max}-E_{min}}\right). \quad (6)$$

where A is normalization constant. Converting $E, E_{min}$ and $E_{max}$ into protein stabilities $\Delta G, \Delta G_{min}, \Delta G_{max}$ using Eqs (1,2) and assuming a standard unfolded state, we get distribution of stabilities of all proteins:

$$p(\Delta G) = e^{-\frac{h\Delta G}{h^2+D}}\sin\left(\pi\frac{\Delta G-\Delta G_{min}}{\Delta G_{max}-\Delta G_{min}}\right) \quad (6a)$$

The population growth rate $\lambda$ is then

$$\lambda = b - d - \frac{1}{2} m\Gamma \left\{ \frac{h^2}{h^2 + D} + \frac{\pi^2 (h^2 + D)}{(E_{max} - E_{min})^2} \right\} \quad (7)$$

The population survives only when $\lambda \geq 0$ which imposes an upper bound $\Gamma^*$ on the number of genes of an organism at a given mutational load:

$$\frac{m\Gamma}{b} < \frac{m\Gamma^*}{b} = \left(1 - \frac{d}{b}\right) \frac{2}{\left\{ \frac{h^2}{h^2 + D} + \frac{\pi^2 (h^2 + D)}{(E_{max} - E_{min})^2} \right\}} \quad (8)$$

Note that $\frac{m\Gamma}{b}$ is simply the number of mutations per portion of the genome encoding essential genes per replication event. Even in the absence of natural death process (i.e. when d=0) the r.h.s. of eq.(8) establishes an absolute upper physical limit on this number for any unicellular, asexually reproducing organism.

Eqs. (6-8) represent the main results of this work and now we turn to their biological implications. First, Eq.(6a) predicts a universal distribution of the stabilities of all existing protein domains  Stabilities of proteins vary in the range from 0 to about 20kcal/mol [15,16], providing an estimate for $\Delta G_{max} - \Delta G_{min} \approx 20 kcal/mol$. As can be seen from Fig.2, Eq.(6a) describes the distribution of protein stabilities quite well with parameters h and D derived from independent point mutation experiments. In particular, it reproduces a characteristic asymmetric distribution of protein stabilities.

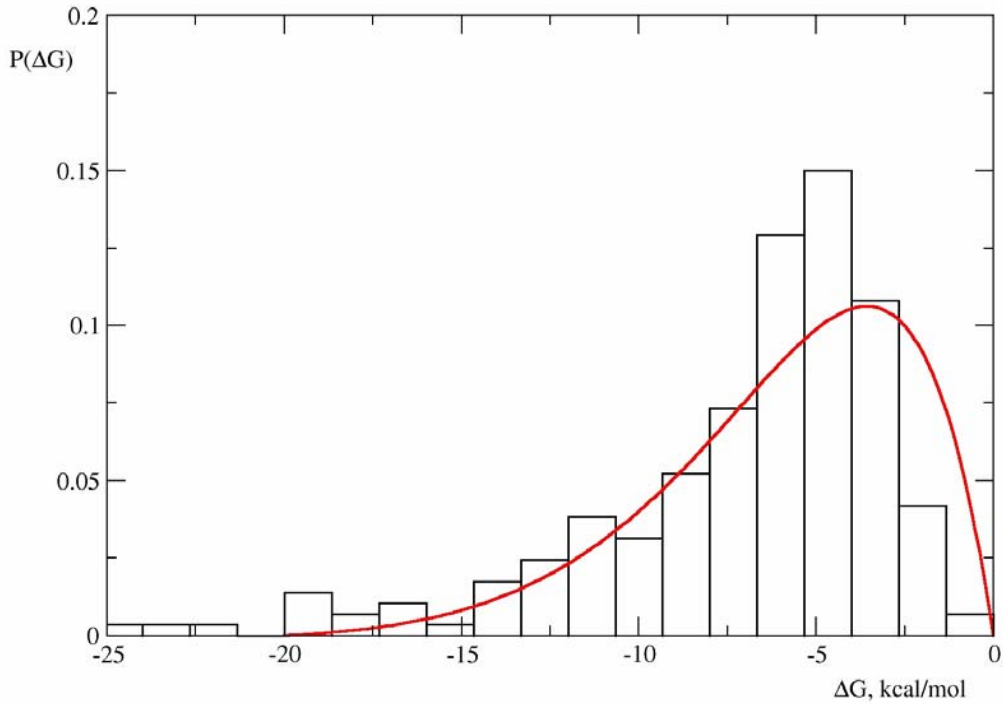

**Figure 2.** *Distribution of stabilities of single domain proteins and the prediction (red line) from analytical equation (6) with parameters h and D derived from mutation data on proteins as explained in the text and Supp Info. The stability data was collected from the ProTherm database [15]. The value $\Delta G_{max} - \Delta G_{min}$ is taken to be 20kcal/mol.*

Eqs.7 and 8 establish an upper limit on the genome sizes of organisms at a given mutational load and replication and death rates. They predict that if the genomes are too large or the mutation rate is too high, populations will go extinct due to lethal mutagenesis. Our theory predicts a universal upper limit threshold of a mutational load that a population can sustain, even without death of organisms or loss of parent genomes due to natural causes (e.g. degradation of parent RNA in viruses). With the estimated values of the parameters h, D, and $E_{max} - E_{min}$, the mutational threshold can be estimated (for d=0) as approximately 6 mutations per essential portion of the genome per replication. This prediction is in good quantitative agreement with many lethal mutagenesis experiments on RNA viruses [17].

Our theory predicts that the maximum essential genome size (number of essential genes) $\Gamma^*$ is smaller for organisms with higher mutation rate $m$. This prediction is in excellent agreement with observations for a broad range of organisms (see Fig.1 in [17], reproduced as Fig. S3 in Supplementary Information). A further example is provided by the data on distribution of viral genome lengths. It is well-known that mutation rate in RNA viruses is much greater than that in dsDNA viruses [17,18]. Correspondingly, the theory predicts much longer genomes for dsDNA viruses than for RNA viruses, in harmony with observations (Fig.3).

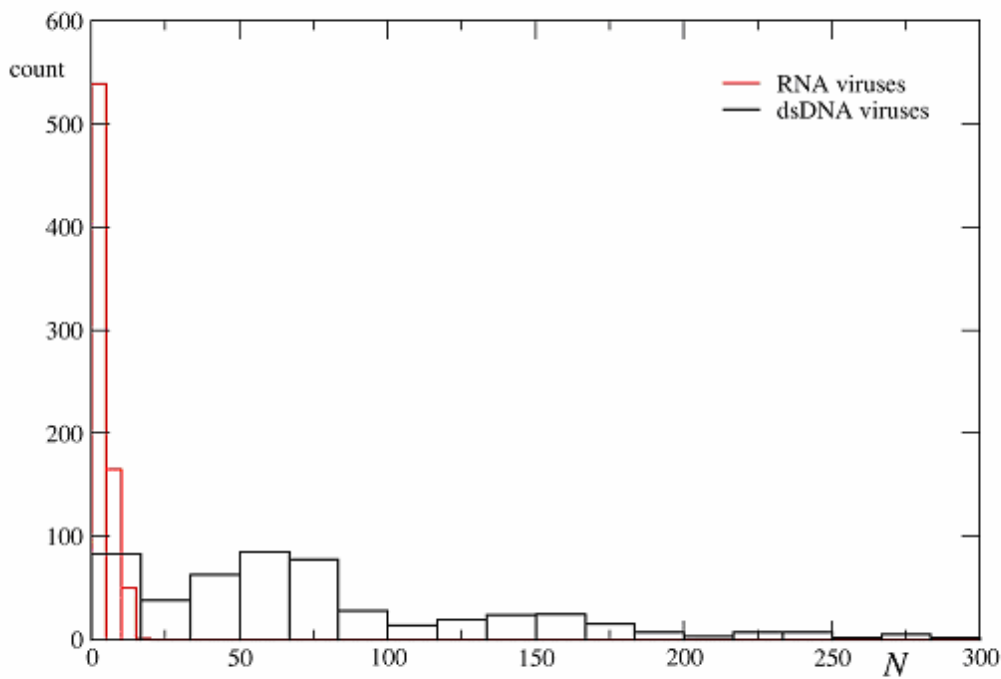

**Figure 3.** *The distribution of number of genes per viral genome. The red histogram corresponds to RNA viruses while black histogram is for dsDNA visruses. The data is taken from NCBI Genome database,*
*http://www.ncbi.nlm.nih.gov/genomes/static/vis.html. The genomes of RNA viruses are much shorter than dsDNA viruses.*

Another important prediction from the theory concerns thermophilic organisms. Stability of proteins at elevated environmental temperature requires that their native state energy should be lower than of mesophilic proteins, according to Eq.(1) (see also Supplementary Information). Correspondingly, the range of possible native state energies $E_{max} - E_{min}$ in Eqs.6-8 shrinks for thermophilic organisms and the mutational meltdown threshold decreases. The magnitude of the effect can be estimated using a typical value of the entropic difference between unfolded and folded states for a typical 100 aminoacid protein domain, 0.25kcal/mol/K [9]. An increase of environmental temperature by 60K, typical of hyperthermophiles results in decrease of the $E_{max} - E_{min}$ from 20kcal/mol to 8kcal/mol, leading to a decrease in the mutational meltdown threshold (r.h.s. of Eq.(8)) - from 6 to 2. The implication of that is twofold. First, it suggests that hyperthermophilic – crenarchaeal – viruses (phages) can be only dsDNA, as RNA viruses are already in the mutational meltdown regime at this temperature. Indeed all known crenarchaeal viruses are dsDNA [19] and so far no extremophilic virus with an RNA genome had been found.

Second, assuming the same mutation rate per nucleotide in mesophilic and thermophilic organisms [20] our theory predicts that organisms living at elevated temperatures should have shorter genomes. In Figure 4, we plotted the number of genes in the genomes of 202 bacteria and archaea (see Table 1 in Supplementary information of [12]) as a function of their optimal growth temperature. It is clearly seen that prokaryotes living at 60°C or above systematically posses shorter genomes (about 2000 genes) than mesophilic prokaryotes with optimal growth temperature of 20-40°C. Existence of mesophilic prokaryotes with short genomes does not contradict our theory, as Eq. (8) sets only an upper limit on the size of the essential proteome; if an organism functions with a smaller number of genes, it simply does not ''feel'' the pressure on the genome size due to protein stability requirements. In these cases, regulation of the genome size is governed by other biological mechanisms, such as the energetic cost of genome duplication and repair.

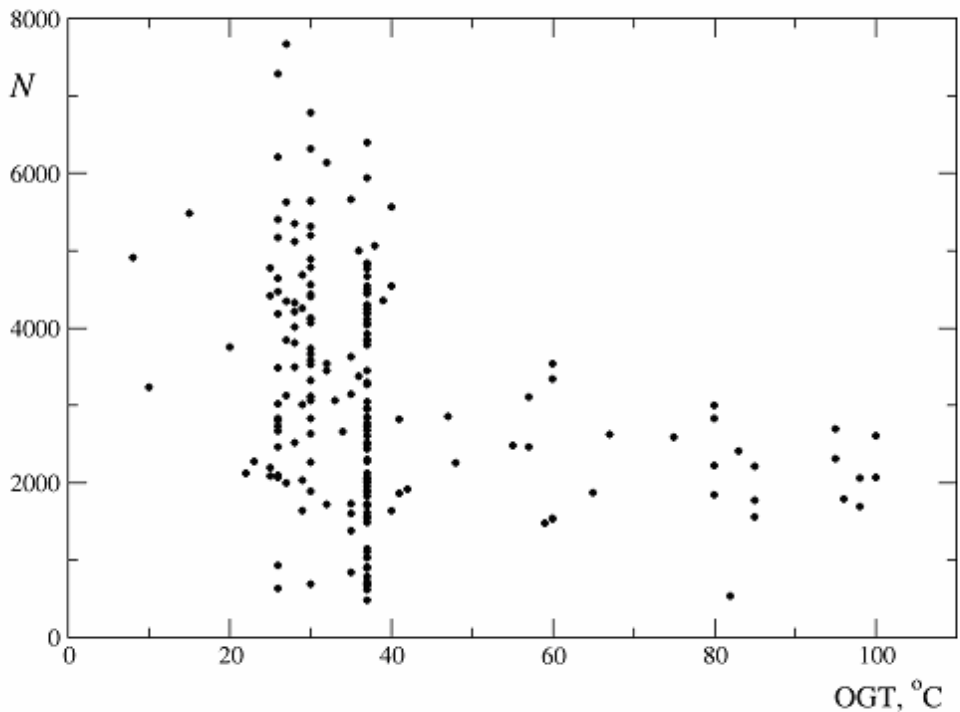

**Figure 4.** *Distribution of the number of genes in 202 prokaryotic genomes for organisms with various optimal growth temperatures. Thermophilic and hyperthermophilic organisms systematically possess shorter genomes than mesophilic ones.*

In contrast to other approaches [6] we did not *a priori* assume here an existence of optimal stability of proteins that renders highest fitness. Such assumptions are often motivated by a circular argument that observed stabilities of proteins are not too high. Here we assume that greater stability of proteins is not functionally advantageous to organisms, as long as proteins possess at least minimal stability ($\Delta G < 0$) to function, On the other hand, sequence entropy factor favors less stable proteins: Sequences that can deliver higher stability become more scarce [8,21]. The entropic factor in sequence space is the main reason of why destabilizing mutations are statistically more frequent (i.e. that mutation drift parameter h in Eq.4 is positive in our theory): it is more likely to find mutated sequences of lower stability than the ones stabilizing a protein [8,21]. (A more detailed analysis that explicitly takes into account sequence statistics, i.e. considers realistic dependence of h and D in Eq.(4) on E [21], supports this conclusion (data not shown)).

One would then expect that most proteins should be ''marginally stable'' with distribution of stabilities peaked close to $\Delta G = 0$ value [7,8]. However, the distribution of stabilities of real proteins has a characteristic peak at a moderate (not too low) value around 5kcal/mol and a sharp asymmetric decrease at both lower and higher stabilities (see Fig.2). The theory explains why naïve expectations of marginal stability of proteins are not borne out. Proteins evolve as part of organisms and very low values of their stability would result in too frequent mutational ''falls from the $\Delta G = 0$ cliff'', resulting in death of the organism and subsequent elimination of the gene encoding the marginally stable protein. This is a clear manifestation of the effect of the constraints imposed by organismal evolution on the distribution of molecular properties of proteins. An important lesson from this study is that a peaked distribution of a specific property of evolved proteins, such as stability $\Delta G$, does not imply or require that proteins with the peak value of the property confer the highest fitness to their carrier organisms. Instead, the peaked distribution can arise from a balance between opposing factors (stability above lethal threshold and sequence entropy) in a locally flat fitness landscape with just two, lethal and viable, phenotypes.

Another key universal prediction of this model is lethal mutagenesis at high mutation rate – about 6 mutations per genome *per replication* for mesophiles and 1-2 for hyperthermophiles living close to 100°C. As this number is per replication, higher replication rate would make it possible to avoid lethal mutagenesis at a given absolute mutation rate. Lethal mutagenesis is often attributed to Eigen's error catastrophe [22][23]. However, this interpretation may be misleading. In fact, at high mutation rates the quasispecies theory predicts a very different phenomenon, delocalization in sequence space. As the quasispecies theory applies only to soft selection, extinction of populations via lethal mutagenesis cannot occur there [24,25]. The lethal mutagenesis predicted in the present work is also different from the Muller ratchet [26] as it can occur even for infinite populations upon exceeding a well-defined mutation rate threshold, in contrast to the Muller ratchet mechanism [27]. As Wilke and coworkers pointed our recently [28] there has been no clear understanding of lethal mutagenesis. Our findings represent a simple first-principles theory of this important phenomenon and may have direct implications for further development of therapies based on mutation-inducing drugs or radiation.

The results presented here are most directly applicable to RNA viruses whose mutation rates are close to the mutational meltdown threshold of Eq.(8). In DNA organisms, mutation rates are typically 2-4 orders of magnitude lower due to the action of error correction mechanisms [18]. However, highly pathogenic strains of bacteria often exhibit mutator phenotypes [29] whose mutation rates can approach the mutational load limit discovered in this work. As in the case of viruses, it was argued that higher mutation rates in some strains of bacteria may emerge to facilitate adaptation to their environment. However, the mutational meltdown threshold puts a physical limit on the mutational response of pathogens to rapidly changing host environments.

Our model is basic, and it does not consider many biologically relevant factors, such as functional selection, epistasis or death of organisms due to environmental fluctuations. Organisms do not interact in the model and do not compete for resources. We consider hard selection whereby population size N is not fixed.

We further simplified our consideration of protein thermodynamics by assuming roughly equal length of all proteins (reflected in universal $E_{max} - E_{min}$ values). It is known that protein lengths vary in a broad range. However, a two-state folding unit to which this theory is applicable is a protein domain. The range of variation of domain sizes is somewhat smaller than that for complete proteins so we made our estimates for a typical 100 aminoacid domain. The variation of domain lengths can be taken into account in further development of the theory.

Our analysis establishes the limits on the genome size that an organism can maintain at a given mutational load. In particular, the limitations set by this theory may be important for understanding the origins of life or *de novo* design of artificial life, as early or newly designed organisms probably could have an imperfect replication machinery, and, thus, elevated mutation rates.

*Acknowledgements.* We are grateful to Boris Shakhnovich, Jesse Bloom and Claus Wilke for many fruitful discussions. This work was supported by the NIH.